# Photoexcitation Induced Quantum Dynamics of Charge Density Wave and Emergence of a Collective Mode in 1*T*-$TaS_2$

Jin Zhang,[1,2] Chao Lian,[1] Mengxue Guan, [1,2,] Wei Ma,[1,2] Huixia Fu,[1,2] and Sheng Meng[1,2,3*]

[1]*Beijing National Laboratory for Condensed Matter Physics, and Institute of Physics, Chinese Academy of Sciences, Beijing 100190, P. R. China*

[2]*School of Physical Sciences, University of Chinese Academy of Sciences, Beijing 100049, P. R. China*

[3]*Collaborative Innovation Center of Quantum Matter, Beijing 100190, P. R. China*

J.Z., C.L. and M.G. contribute equally to this work.

*Corresponding author: Sheng Meng (smeng@iphy.ac.cn)

# Abstract

**Photoexcitation is a powerful means in distinguishing different interactions and manipulating the states of matter, especially in complex quantum systems. As a well-known charge density wave (CDW) material, 1*T*-$TaS_2$ has been widely studied experimentally thanks to its intriguing photoexcited responses. However, the microscopic atomic dynamics and underlying mechanism are still under debate. Here, we demonstrate photoexcitation induced ultrafast dynamics in 1*T*-$TaS_2$ using time-dependent density functional theory molecular dynamics. We discover a novel collective mode induced by photodoping, which is significantly different from thermally-induced phonon mode in $TaS_2$. In addition, our finding validates nonthermal melting of CDW induced at low light intensities, supporting that conventional hot electron model is inadequate to explain photoinduced dynamics. Our results provide a deep insight on coherent electron and lattice quantum dynamics during the formation and excitation of CDW in 1*T*-$TaS_2$.**

***Key words*** ***：*** *Charge density wave, TDDFT, Photoinduced Phase transition, 2D Materials* *，*

## Introduction

Interplay among different degrees of freedom including electrons, phonons and spins is of paramount importance in understanding and optimizing the properties of quantum materials.[1-3] Optical excitation is a powerful tool to distinguish different interactions and to manipulate the state of matter. Thus, dominant interactions can be identified and meaningful insights into ground-state properties, phase transitions, and hidden phases can be obtained.[4-8] It is particularly useful for complex quantum systems, where a variety of degrees of freedom and quantum interactions coexist and are strongly coupled.

A particular example is charge density wave (CDW) materials.[9-16] The layered transition-metal dichalcogenides such as 1*T*-$TaS_2$ have been widely investigated to understand CDW physics in real materials.[17-26] 1*T*-$TaS_2$ is a typical quasi two-dimensional CDW material with a pristine lattice constant of 3.36 Å in undistorted 1*T* phase (referred to as T state). The canonical origin of CDW phase is usually attributed to the Fermi-surface nesting, driven by electron-phonon coupling via the Peierls mechanism.[27-30] In the low-temperature commensurate CDW phase, referred to as C state hereafter, 1*T*-$TaS_2$ becomes strongly modulated, leading to a lattice distortion with a new periodicity of ~12.1 Å rotated by 13.9°. Charge transfer from the outer ring towards the center leads to the famous pattern, namely, the “star of David” (SD) shape of the lattice distribution.

Laser induced phase dynamics in 1*T*-$TaS_2$ has been widely investigated experimentally in recent years.[17,21-26] Excitation with ultrashort laser pulses closes both CDW and Mott gaps in 1*T*-$TaS_2$. At low intensities it also excites the breathing phonon mode, *e.g.* amplitude mode of the CDW, where the atomic clusters and the charge order breathe synchronously around the equilibrium state.[21-23,25-26] The ultrafast melting of Mott gap upon strong laser excitation has been observed, while the lattice retains its low-temperature symmetry.[23] However, it is elusive whether the lattice degree of freedom and ionic movements are *simultaneously* coupled to electronic modulations in this case, and the underlying mechanism of photoexcitation under extreme conditions is not clear.[21-23]

To explain phase dynamics of 1*T*-$TaS_2$ under laser illumination, hot electron model was widely invoked to understand the interactions between electron and lattice subsystem.[17,21-22] In such a model, the excited electrons rapidly thermalize among themselves forming hot electron gas with a given electronic temperature ($T_e$) of several thousands of Kelvin, significantly higher than that of the lattice subsystem. The cold lattice is then heated up by hot electrons via effective electron-phonon interactions. The melting of CDW state and band-gap closing in turn takes places when the lattice temperature is higher than its equilibrium thermal melting temperature. Nevertheless, it conflicts with the fact that excited electrons and the lattice take a quite long time (>1 ps) to reach equilibrium with a Fermi-Dirac electronic distribution[31], while in experiment an ultrafast (~30 fs) band-gap closing was observed[23]. Another physical picture is that phase dynamics is induced by the modification of the potential energy surfaces. Strong lasers and ultrafast electron-electron scattering raise the electronic temperatures to several thousand Kelvin, subsequently modifying the ground-state potential surface. Furthermore, it leads to cooperative atomic motions towards a new photoinduced phase, which is usually considered to be the undistorted 1*T* phase of bulk $TaS_2$.[21] However, the microscopic atomic dynamics and underlying mechanism for 1*T*-$TaS_2$ under laser excitations are still under debate.

In this work, we investigate the atomistic mechanism and ultrafast photoinduced dynamics of CDW in 1*T*-$TaS_2$, using nonadiabatic molecular dynamics (MD) simulations based on time-dependent density functional theory (TDDFT). Our first-principles simulations yield for the first time intrinsic electron-nuclei coupled dynamics of 1*T*-$TaS_2$.[32-39] Amplitude mode and nonthermal melting of CDW are successfully reproduced at low laser intensities. At higher laser intensity a laser-induced new collective mode is resulted, with distinctive electronic properties. This work not only gives a deep insight into the photoinduced nonequilibrium dynamics of 1*T*-$TaS_2$, but also provides a framework to understand laser induced phenomena in more quantum materials.

## Results

**Systems and thermal melting.** At high temperatures 1*T*-$TaS_2$ is metallic, exhibiting an undistorted lattice. Below 350 K, the SD shaped patterns of tantalum atoms show up, accompanied with a periodical lattice distortion. At temperatures below 180 K, the CDW state brings an insulating electronic structure, due to the presence of a fully commensurate CDW with the long-ranged $\sqrt{13} \times \sqrt{13}$ superlattice[20]. Figure 1(a) shows the atomic structure of 1*T*-$TaS_2$ in its ground-state C phase with the SD pattern, which is energetically favored over the T phase (by 73 meV/atom). A band gap of 0.45 eV in CDW state of $TaS_2$ is induced by periodical in-plane lattice distortion, consistent with previous experimental and theoretical studies.[21,40-41]

We focus on the dynamics of CDW phase under different thermal and laser conditions. To demonstrate the behavior of bulk 1*T*-$TaS_2$ under different temperatures, root mean square displacement (RMSD) of all atoms is utilized to quantitatively analyze lattice structural changes. Figure 1(b) shows the evolution of RMSD under different ionic temperatures in Born-Oppenheimer molecular dynamics (BOMD). The equilibrium value of RMSD increases as the ionic temperature rises. We find that the critical point to melt CDW state is located at ~400 K, when the RMSD reaches 0.2 Å. At this point, the ionic SDs start to disappear and the lattice restores its undistorted 1*T* geometry where the spatial modulation becomes much weaker. [Snapshots of atomic structures are presented in Figure 1(c) and more details can be found in the Supplemental Information (SI)]. We adopt a criterion that the CDW state melts or transforms to a new phase when the RMSD value reaches $R_c$ = 0.2 Å, based on the collapse temperature of the SDs. The experimentally observed equilibrium distortion of CDW state is 7% of in-plane lattice constant, equaling to ~0.24 Å (see SI), which rationalizes the value obtained here.[20] Thus, the CDW state in bulk 1*T*-$TaS_2$ melts thermally at about 400 K with a maximum RMSD of 0.2 Å in our simulations. The small deviation from the experimental value of 350 K may arise from the supercell constraint of the model used. We note that the nearly commensurate and incommensurate CDW phases cannot be directly simulated because of periodic boundary conditions and small supercells used here.

**Time-dependent evolution.** To illustrate laser-induced charge modulation in bulk 1*T*-$TaS_2$, we elaborately build initial states of photoexcitation by changing the population of Kohn-Sham orbitals from ground state to specific configurations, according to the calculated matrix elements of optical transitions at ~1.6 eV (see SM). Energy distribution of excited electrons/holes agrees well with time-dependent simulations, suggesting nonlinear light absorption is not significant (see SI). Here we use the percentage of valence electrons pumped to conduction bands to denote the laser intensity $\eta$. For instance, $\eta$ = 1% means 1% of total valence electrons are promoted to specified unoccupied bands (laser-induced charge density differences are shown in SI). Optical excitation will modify the occupation of electronic orbitals and the potential energy between ions, thus the total energy ($E_{laser}$) would change significantly. The difference between $E_{laser}$ and the ground-state energy ($E_0$) gives the excitation energy per atom

$$E_{ex} = (E_0 - E_{laser}) / N_{atom} \ ,$$

where $N_{atom}$ is the number of the atoms in the simulation. We estimated the fluence of applied laser based on the calculated excitation energy. For $\eta$ = 0.64% with the calculated $E_{ex}$= 40.5 meV/atom, the laser intensity is estimated to be 0.25 mJ/cm$^2$ (assuming about 10% of the laser energy is absorbed). We find that optical excitations bring about an increase (yellow contour) in charge density near the center of the SD and a clear depletion (light blue contour) around the edge of the stars.

Next, we consider ultrafast photoinduced dynamics in the bulk 1*T*-$TaS_2$ structure under different laser intensities. At low laser intensity $\eta$ = 0.64% (ca. 0.25 mJ/cm$^2$, corresponding to the excitation energy of 40.5 meV/atom or 0.11 excited electrons/$TaS_2$), atomic positions of 1*T*-$TaS_2$ change very little [Figure 2(a)]. The RMSD increases to only 0.06 Å after about 200 fs and then stays stable at such a value [black line in Figure 2(b)]. We note that a small modulation in RMSD shows up with a period of ~400 fs, corresponding to the amplitude mode of CDW.

For a larger laser intensity $\eta$ = 1.28% (ca. 0.5 mJ/cm$^2$), we observe a phase destruction (melting) of the SD at ~250 fs after photoexcitation in Figure 2. The value of RMSD reaches a maximum (0.25 Å) at 440 fs, larger than the critical value of melting ($R_c$ = 0.2 Å). We monitor the corresponding lattice temperatures after the photoexcitation in Figure 2(c) and find the temperature grows to 103 K after 100 fs, attributed to the energy transfer from Kohn-Sham electronic orbitals to the kinetic energy of the lattice subsystem. This can be regarded as the timescale for excited electrons and holes transfer their excess energy to the lattice, *e.g.*, the timescale of electron-phonon coupling. The concurrent ionic temperature is evidently lower than that of thermal melting (~400 K), indicating that the ultrafast CDW melting is not induced by lattice temperature higher than the melting point in equilibrium.

For an even larger laser intensity of $\eta$ = 1.92% (ca. 0.75 mJ/cm$^2$), laser induced phase dynamics in bulk 1$T$-$TaS_2$ is also shown in Figure 2. Very surprisingly, the electron-nuclear dynamics is very distinctive from above cases: not only the RMSD exhibits a periodical oscillation with a time period of 480 fs, but also a new photoinduced transient metallic state (referred to as M state) with spatially-ordered atomic structures is obtained. Eventually, the maximum of RMSD reaches 0.32 Å at 250 fs in Figure 2(b), where the SD collapses evidently and a new order is generated. After that the value of RMSD decreases to 0.1 Å below $R_c$ and the atomic structure resembles that in C state, which is fully restored at $t$ = 480 fs. In the meantime, the corresponding ionic temperature grows to 270 K within 110 fs and then oscillates around 200 K [red line in Figure 2(c)]. For larger laser intensity of $\eta$ = 3.2% (ca. 1.25 mJ/cm$^2$), the features of charge-lattice dynamics is similar, despite that the lattice temperature reaches above 400 K (see SI). The photoinduced transient state also appears at 210 fs with a maximum RMSD of 0.35 Å. Therefore, larger laser intensity will not change the charge-lattice correlated dynamics, except that the equilibrium ionic temperature of the system increases to ~400 K. Surprisingly, no melting of CDW order is observed for laser intensities above $\eta$ = 1.92%, even at a high lattice temperature > 400 K.

It should be noted that the lattice temperature [Figure 2(c)] in a laser-induced nonequilibrium system ($\widetilde{\mathrm{T}}$) is distinct from the equilibrium ionic temperature (T). The $\widetilde{\mathrm{T}}$ comes from the kinetic energy of all ions defined as: $\tilde{T} = Mv_{ions}^2/3k_B$, where $v_{ions}$ is the ion velocity, $M$ is the atomic mass and $k_B$ is the Boltzmann constant. The large oscillations in $\widetilde{T}$ can be interpreted as that the kinetic energy and ionic potential energy exchanges coherently, while the small derivations at the later stage (*e.g.*, at 400 K) may be due to fluctuations. Our TDDFT-MD simulations naturally include the damping of high-energy quasiparticles, which are dissipated to electrons at the lower levels and to the ionic subsystem as shown in Figure 2.

Interestingly, the experimentally observed periodical oscillation has been successfully reproduced in our simulations (Figure 3).[8,25] The trends and features of the experimental data and our simulated X-ray diffraction intensity are almost identical, demonstrating our simulations capture the most important interactions in experiment.[8] The calculated excitation energy $E_{ex}$= 141.2 meV/atom and laser fluence 0.75 mJ/cm$^2$ is in good agreement with experimental data (0.56 mJ/cm$^2$).[20,25] In addition, vibrational analysis shows that the photoinduced atomic displacements are quite different from that of the thermally-induced phonon mode in the ground-state C phase, which has a different vibration pattern and an oscillation frequency of ~2.3 THz (corresponding to a period of 435 fs) [21-22]. It is interpreted that the intriguing collective mode is attributed to the large modulation in the potential energy surface (see SI).

We notice some experimental works reporting on photoexcitation induced "hidden" metastable state or domain-like ordered state.[13,43-47] For instance, Han *et al.* investigated optical doping-induced transition states and phase diagram in such a complex materials, providing robust evidences of optically-driven nonthermal phase transition in 1*T*-$TaS_2$.[47] However, the problem solved there is different from the one we discussed: we reveal the initial electronic and atomic dynamics of 1*T*-$TaS_2$ at the first several hundred femtoseconds, which is also the focus in many recent experimental publications[20-25]. Photoinduced atomic dynamics studied here only lasts < 1 ps, too short to generate any ordered domain. The same short-time dynamics was also observed at the first few picoseconds in photoinduced domain-like states[13]. Therefore, our result is an important support and supplement to fully understand photoinduced processes observed in these experiments.

**Photoinduced transient state.** To obtain more information about the photoexcitation induced M state [Figure 4(a)], we compare its structural characteristics with that of T state without any atomic distortion and the low-temperature C state. It is evident that three peaks (3.20 Å, 3.60 Å, 3.90 Å) dominate in the distribution of the Ta-Ta distances in C state [Figure 4(c)], while the distribution of the nearest Ta-Ta distances in the new M state [blue envelope line in the bottom of Figure 4(c)] exhibits only two peaks (3.36 Å and 3.75 Å). It should be noted that the Ta-Ta distances have the uniform value of 3.36 Å in original undistorted T state. The new emerging peak in M state around 3.75 Å is attributed to the Ta-Ta distances between the atoms in the regular central hexagons and those in the outer rings, indicating the photo-induced state is far from the T phase.

In addition to the difference in atomic structures, electron densities of states of three phases (T, C and M) are also different as shown in Figure 4(d). We obtain quite a few states around Fermi level for M phase, which is in obvious contrast with the bandgap for C state with SD. Despite both T and M states are metallic, the electronic properties are different: T state has a sharper dispersion at Fermi level compared to M state. As a result, the features in optical absorption for T and M states are also different (see SI). Experimentally, laser pulse with a photon energy around 1.6 eV can be used to excite abundant electrons from the valence bands to conduction bands in bulk 1*T*-$TaS_2$ at low temperature to test our prediction and the laser strength of $\eta$ = 1.92% can be easily realized[21,42]. Furthermore, the new emerging M state may be verified by electron or X-ray diffraction analysis, and tested against the simulated diffraction patterns shown in Figure 4(b). We emphasize that the photoexcitation-induced M state cannot be reached *thermally*, where the lattice displacement is at most 0.2 Å before the SD pattern melts into randomized T state at a high temperature > 400 K.

Besides the photoinduced cooperative atomic motions in bulk 1*T*-$TaS_2$, useful information is also obtained from the evolution of orbital energies during the photoexcitation electron-lattice dynamics [see SI]. For a low laser intensity ($\eta$ = 0.64%) the energy gap is preserved at all times, while the gap closes in a subvibrational timescale (~50 fs) for $\eta$ = 1.92%, indicating the electron modulation in valence bands are suppressed and charge order unlocked to the lattice before the atomic structure responds to laser illumination. These results are in good accordance with experimental observations[22], where not only the Mott gap at the Fermi level but also the CDW gap melts on subvibrational timescales, confirming electron-electron scatterings is at the heart of the dynamics of CDW.

## Discussion

Our calculations reveal that LDA/GGA without adding additional *U* describes the formation and dynamics of CDW reasonably well. However, the effect of Mott insulating nature should be paid more attention when we consider *monolayer* 1*T*-$TaS_2$ or very thin films. For monolayers, we note that onsite Coulomb interactions of electrons localized within the SD cause a band separation at the Fermi level by the so-called Mott gap, with a value of 0.20 eV[40,41], while the onsite interaction for the bulk phase plays a negligible role in the band structure [see SI].

We propose that a new intriguing collective mode and a metallic transient state with distinct geometry and electronic properties can be introduced by photoexcitation, which cannot be reached by thermal phase transitions. The obtained phase diagrams of bulk 1*T*-$TaS_2$ under different temperatures and laser intensities are summarized in Figure 5. The low-temperature C state melts at about 400 K [Figure 5(a)] due to the thermal activation, where the average equilibrium velocity of Ta is $1.4\times10^{-3}$ Å/fs at the melting point. For the laser pulse with $\eta$ = 1.28%, C state transforms to the undistorted *T* state, since the laser induced velocity resulted from the changed potential energy surface is about $1.2\times10^{-3}$ Å/fs, very close to the thermal velocity of $1.4\times10^{-3}$ Å/fs at 400 K, resulting in a disordered distribution of Ta ions by both laser and thermal activation. However, at $\eta$ > 1.92%, analysis of atomic trajectories shows that the velocities of Ta atoms in the outer ring ($4.4\times10^{-3}$ Å/fs) are significantly larger than thermal velocities of $1.4\times10^{-3}$ Å/fs, with a highly inhomogeneous and anisotropic distribution over Ta atoms in the inner and outer rings, thanks to dramatically modified potential energy surfaces upon photoexcitation. The emergence of the new collective mode comes from the laser-driven dramatic changes in the interacting potential between Ta atoms. As noted, we find that upon the immediate strong excitation, the laser-induced forces on the outmost Ta in the SD is the largest (~1.4 eV/Å), while the forces acting on other Ta ions remain a relative small value (< 0.6 eV/Å). The forces are ordered and very different from the distributions in thermal equilibrium. This means that stronger laser intensities lead to specific atomic oscillation modes, which strongly suppress thermal movements of Ta and S atoms and the melting of CDW ordering. The transition from $\eta$ = 1.28% to $\eta$ = 1.92% in Figure 5(b) is thus attributed to the competition between temperature-dependent thermal movements and laser-induced lattice oscillations.

In order to find out whether the dynamics is provoked by hot electrons or by excited electrons in nonequilibrium distribution[20], we also performed TDDFT calculations with different initial electronic temperatures from 3000 K to 5000 K with an initial ionic temperature of 10 K. For all electronic temperatures, the ionic temperatures increase at the first 100 fs and RMSDs follow the similar trend, where no periodical oscillations are observed (see SI). More importantly, there is no collapse of electronic band gap. When the excitation energy from physical photoexcitations discussed above and that from the hot electron model are compared (Table 1), it is clear that hot electron model requires a higher energy, while leading to no melting of CDW nor the bandgap closing. This confirms that hot electron model with a well-defined electronic temperature overestimates photoexcitation energies, and electron-electron scattering is vital in the photodynamics of 1*T*-$TaS_2$. Notably, it is not clear whether Fermi surface nesting plays a significant role for the formation of CDW.[50-51]

The direct structural and electronic dynamic data obtained from first-principles calculations enable us to elucidate the photoinduced dynamics after optical excitations of 1*T*-$TaS_2$ at the atomistic spatial scale and femtosecond timescale. We deduce that excited electrons, which do not reach an equilibrium state with a well-defined electronic temperature, play a crucial role in the dynamics of CDW state. Strong photoexcitation creates a high density of electron-hole pairs and the charge order collapses in less than 50 fs before the lattice responds, due to electron-electron scattering. Then electron-phonon interactions raise the effective lattice temperatures from 10 K to 275 K in about 150 fs. Before this takes place, excited electrons strongly modulate the potential energy surface. Thus, it brings up cooperative atomic motions towards a new transient M state far from thermally induced T state and also provokes a novel oscillation mode with a periodicity of ~480 fs, attributed to the energy transfer from electronic subsystem to lattice.

In conclusion, *ab initio* TDDFT-MD simulations reveal the nature of photoexcitation induced phase dynamics in bulk 1*T*-$TaS_2$. We discover a novel collective mode induced by photodoping, which is significantly different from thermally-induced phonon mode in 1*T*-$TaS_2$. Our results provide compelling evidences that the ultrafast dynamics in CDW state of bulk 1*T*-$TaS_2$ is a nonthermal process, where hot electron model is not sufficient to describe this novel phenomenon because of the lack of electron-electron scatterings. Our work yields new insights into laser-induced insulator-to-metal transition in the CDW state of 1*T*-$TaS_2$, and the methods adopted here might be useful for understanding a wide range of laser-modulated quantum materials.

## Methods

**Evolution of electron-ion system.** The nonadiabatic dynamics calculations are performed with a real time TDDFT code, time dependent ab initio package (TDAP)32-35 as implemented in SIESTA36-38. The bulk 1T-TaS2 in its C state is simulated with a supercell of 78 atoms with√13×√13×2 periodical boundary conditions. The Troullier-Martin pseudopotentials and the adiabatic local density approximation for the exchange-correlation functional39 are used. An auxiliary real-space grid equivalent to a plane-wave cutoff of 200 Ry is adopted. During dynamic simulations the evolving time step is set to 50 as for both electrons and ions in a micro-canonical ensemble. As the ions are much heavier than electrons by at least three orders of magnitude, the nuclear positions are updated following the Newton's second law:

$$M_J \frac{d^2 R_J(t)}{dt^2} = -\nabla_{R_J}\left[V_{\text{ext}}^J\left(R_J,t\right) - \int \frac{Z_J \rho\left(r,t\right)}{\left|R_J - r\right|} dr + \sum_{I \neq J} \frac{Z_I Z_J}{\left|R_J - R_I\right|}\right] \quad (1).$$

where, $M_J$ and $R_J$ are the mass and position of the *J*th ion, respectively.

Note that we ignore the negligible ion-ion exchange-correlation function and assume that the ionic density has a sharp distribution $\rho_J\left(R,t\right) = \delta\left(R - R_J\left(t\right)\right)$. The motions of electrons follow the time-dependent Kohn-Sham equations:

$$i\hbar \frac{\partial \phi_j\left(r,t\right)}{\partial t} = \left[-\frac{\hbar^2}{2m}\nabla_r^2 + v_{\text{ext}}\left(r,t\right) + \int \frac{\rho\left(r',t\right)}{\left|r - r'\right|} dr' - \sum_i \frac{Z_J}{\left|r - R_J\right|} + v_{\text{xc}}\left[\rho\right]\left(r,t\right)\right]\phi_j\left(r,t\right) \quad (2).$$

Equation (1) and (2) represents the time-dependent coupled electron-ion motion. The time-dependent Kohn-Sham equations of electrons and the Newtonian motion of ions are solved simultaneously, with ionic forces along the classical trajectory evaluated through the Ehrenfest theorem.

**Preparation of electronic excited states.** To mimic photoexcitation of different laser pulses, we elaborately build the initial electronic states upon photoexcitation. Here, we change the population of Kohn-Sham orbitals from the ground state to specific electronic configurations. The population variation is proportional to optical transition probability. Laser-induced changes in electron density $\Delta\rho(t_0)$ can be expressed as

$$\Delta\rho(t_0) \propto \sum_{<i,j>} P_{ij}[|\psi_i(t_0)|^2 - |\psi_j(t_0)|^2] \quad (3).$$

The transition probability $P_{ij}$ from the initial state $|\psi_i>$ to the final state $|\psi_j>$ fulfills Fermi's golden rule,

$$P_{ij} = |<\psi_i|e\cdot P|\psi_j>|^2 \, \delta(\varepsilon_i - \varepsilon_j - \hbar\omega_l) \quad (4),$$

where $<\psi_i|e\cdot P|\psi_j>$ is the transition matrix element, ***e*** is the amplitude of the electric field, **P** is the momentum operator, $\omega_l$ is the laser frequency, and $\varepsilon_i$ and $\varepsilon_j$ are the energies of the state $|\psi_i>$ and $|\psi_j>$, respectively. See SI for the comparison of different methods to prepare the initial excited states.

**Born-Oppenheimer MD.** The structure optimization and Born-Oppenheimer MD for the ground state were performed at the Γ point in a large supercell with 78 atoms ($\sqrt{13} \times \sqrt{13} \times 2$). After geometry optimization at 0 K, the bulk 1*T*-$TaS_2$ is heated to different temperatures (100 to 500 K) for 1 ps to obtain the thermal dynamics. During dynamic simulations, the evolving timestep is set to 1 fs for both electrons and ions in a micro-canonical ensemble.

**Acknowledgements**

This work was supported by National Key Research and Development Program of China (Grant Nos. 2016YFA0300902 and 2015CB921001) and National Natural Science Foundation of China (Grant Nos. 11474328 and 11290164).

**Author contributions**

S.M. and J.Z. designed the research. Most of the calculations were performed by J.Z., with contributions from all authors. C.L. implemented a time-dependent density functional theory into the SIETSA code. J.Z. and C.L. contributed equally to this work. All authors contributed to the analysis and discussion of the data and the writing of the manuscript.

**SUPPLEMENTARY MATERIALS**

## Figures and Captions

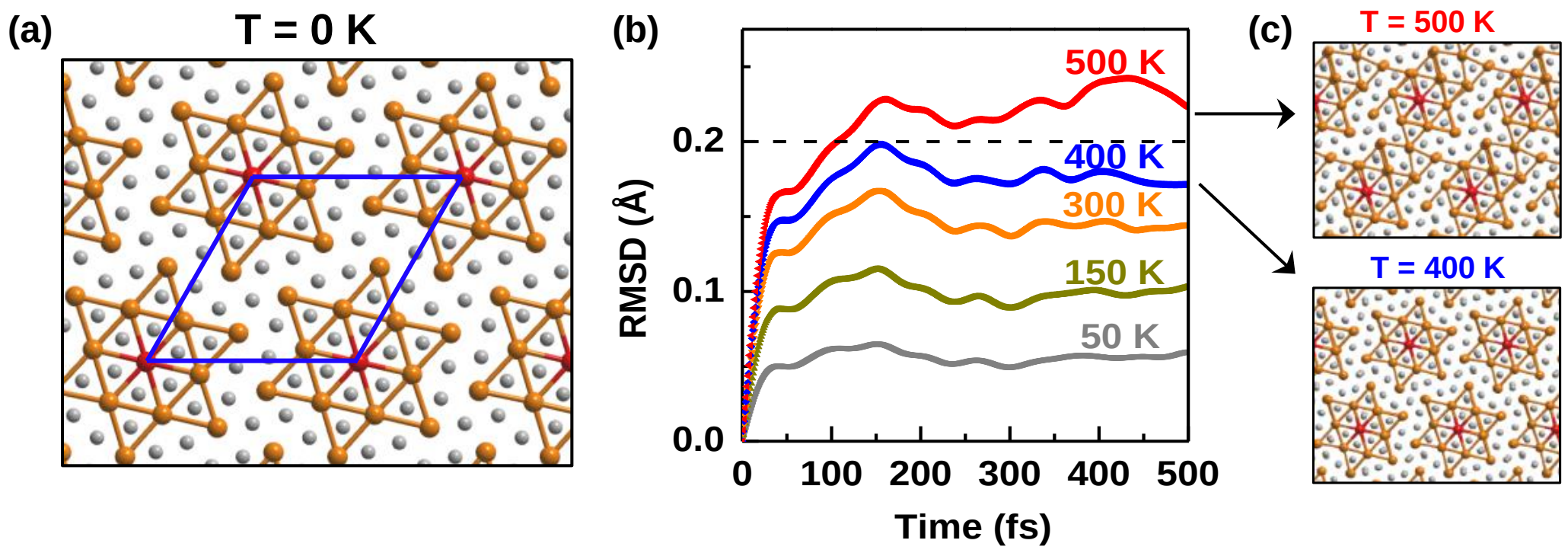

**Figure 1 | Charge density wave state of bulk 1*T*-TaS2 and its thermal melting.** (a) Atomic structure of bulk 1*T*-$TaS_2$ in C state with a $\sqrt{13} \times \sqrt{13}$ superlattice (blue rhombus). (b) Root mean square displacement (RMSD) of 1*T*-$TaS_2$ under different ionic temperatures from Born-Oppenheimer molecular dynamics (BOMD) simulations. (c) Snapshots of the structures at 500 fs of BOMD at 400 K (lower panel) and 500 K (upper panel). The dashed line indicates the critical value for destruction of periodically modulated structures. Orange and gray spheres denote Ta and S atoms, respectively, and red spheres are Ta atoms at the center of stars. The criterion for making Ta-Ta bonds is that the Ta-Ta distance is shorter than 3.4 Å.

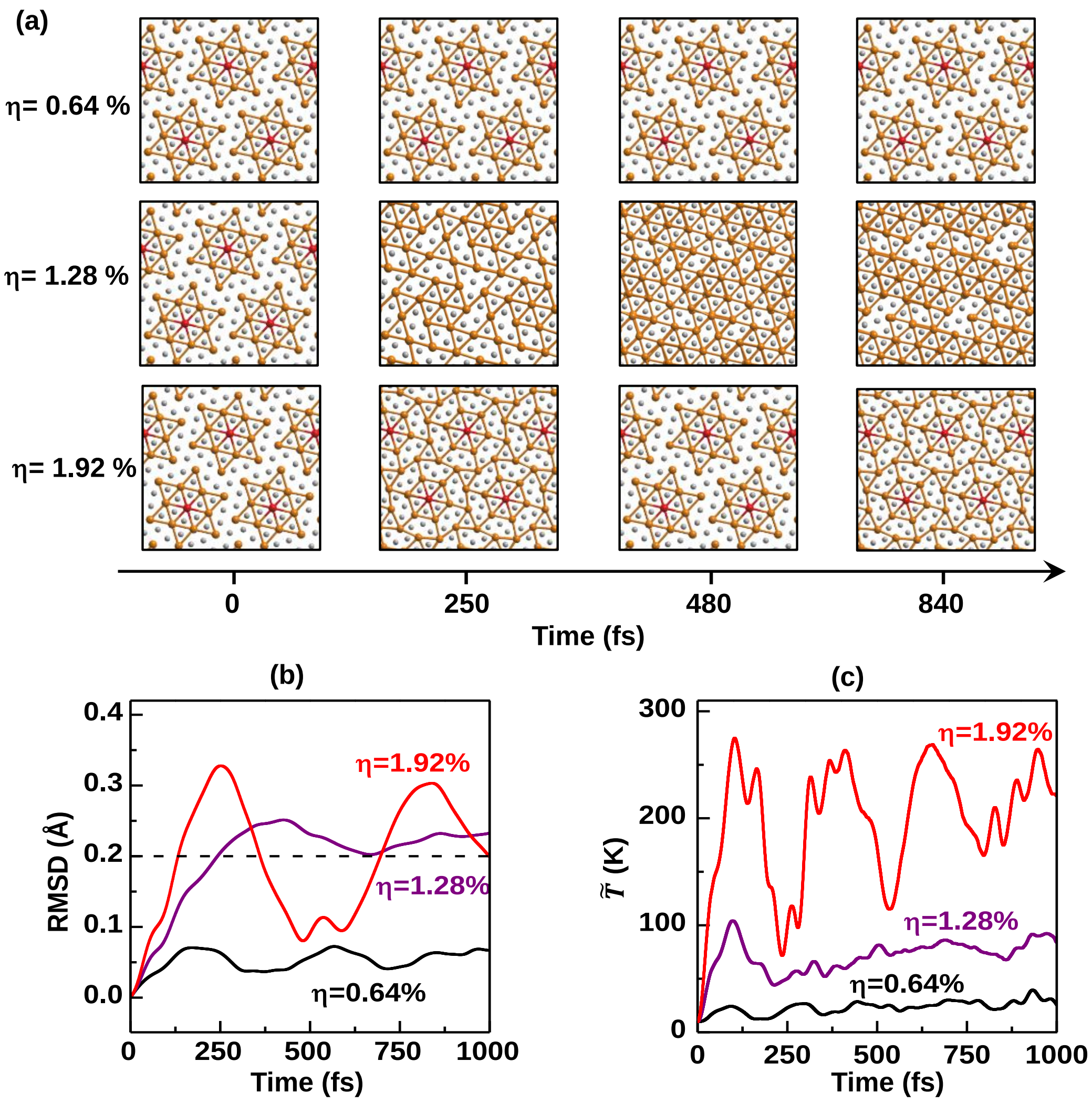

**Figure 2 | Time evolution of atomic structures of bulk 1*T*-$TaS_2$ under different photoexcitation.** (a) Snapshots of time-dependent atomic structures for $\eta$ = 0.64%, $\eta$ = 1.28% and $\eta$ = 1.92% at 0, 250, 480 and 840 fs after photoexcitation, respectively. (b) The evolution of RMSD under three laser intensities (black line: $\eta$ = 0.64%, purple line: $\eta$ = 1.28% and red line: $\eta$ = 1.92%). $R_c$ is shown in dashed black line. (c) Corresponding evolution of the ionic temperatures calculated from the kinetic energy of all ions at different times.

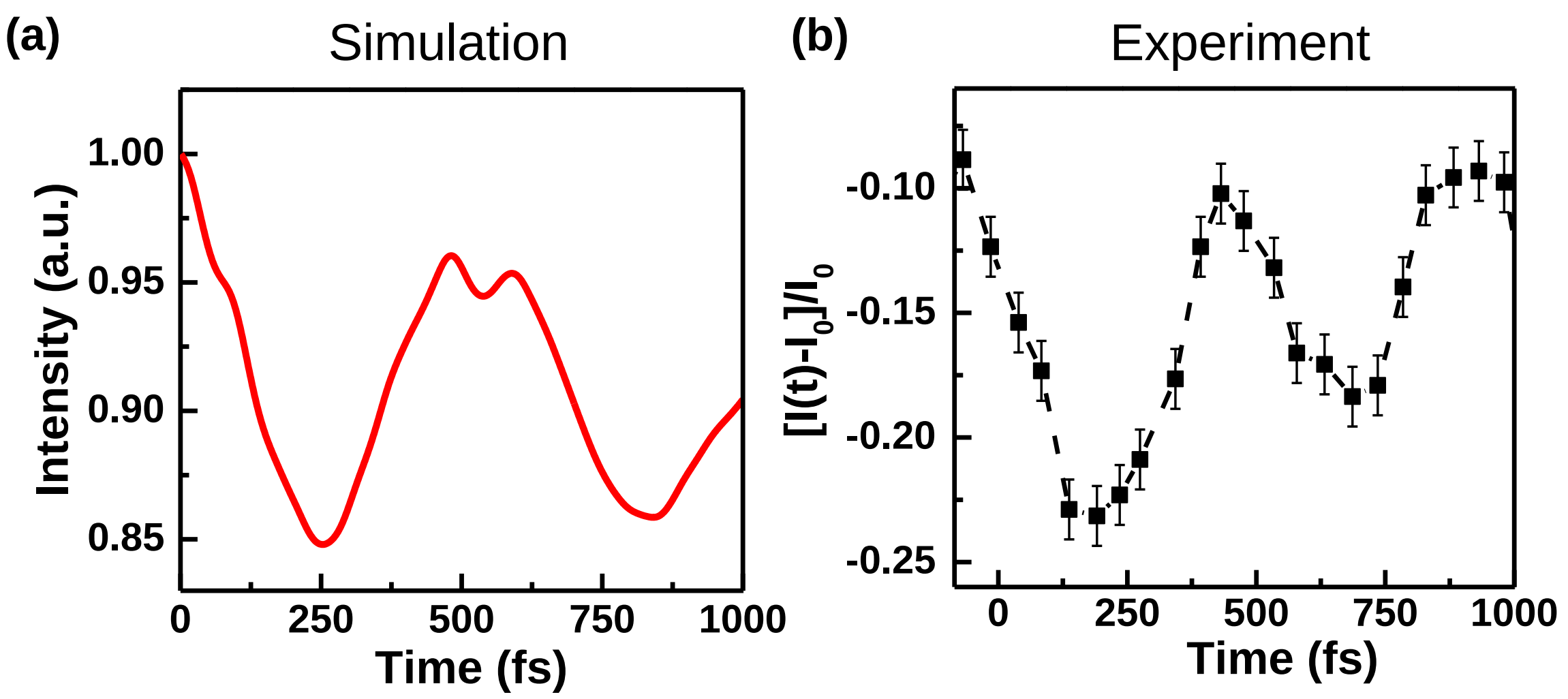

**Figure 3 | Simulated and experimental X-ray diffraction intensity.** (a) Simulated X-ray diffraction intensity as a function of time for $\eta$ = 1.92%. The RMSD is related to diffraction intensity $I(t)$ through the Debye-Waller formula, $I(t)= \exp[-Q^2 \langle u^2(t)\rangle/3]$, where $Q$ is the reciprocal lattice vector of the probed reflection, $u^2(t)$ the square of RMSD. (b) Relative change of diffracted intensity at the satellite position (1.307, 1.231, 0) as a function of pump-probe delay in bulk 1*T*-$TaS_2$ from Ref 8.

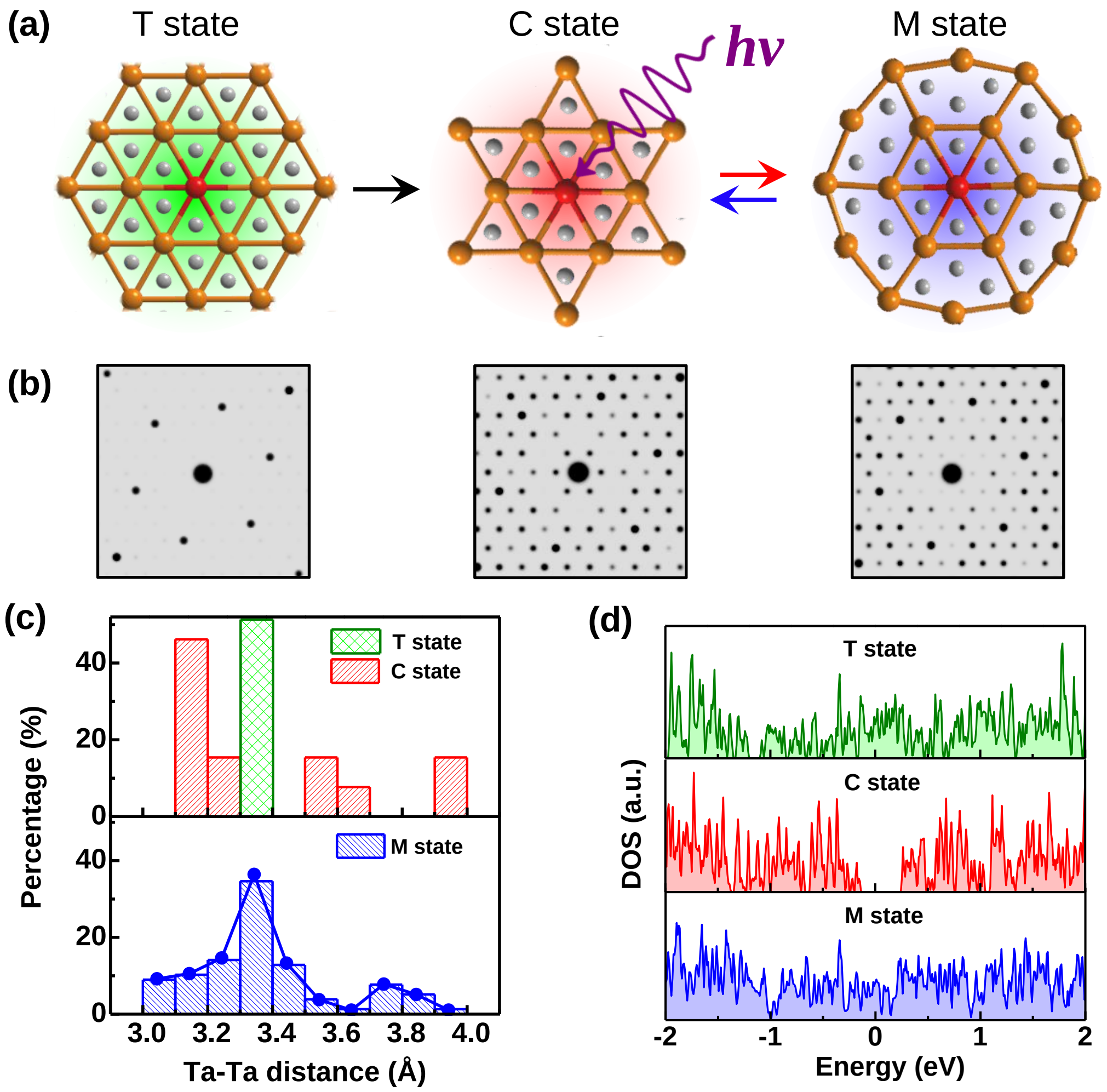

**Figure 4 | Atomic and electronic properties of the three states in bulk 1*T*-TaS₂.** (a) Schematic for transformation between undistorted T state (left), C state (middle) and photoinduced M state (right). (b) Simulated electron diffraction patterns for the three states. (c) Radial distribution of Ta-Ta distances in different states. The T phase is presented in green with the nearest Ta-Ta distances of 3.36 Å. (d) In-plane density of states for the three states.

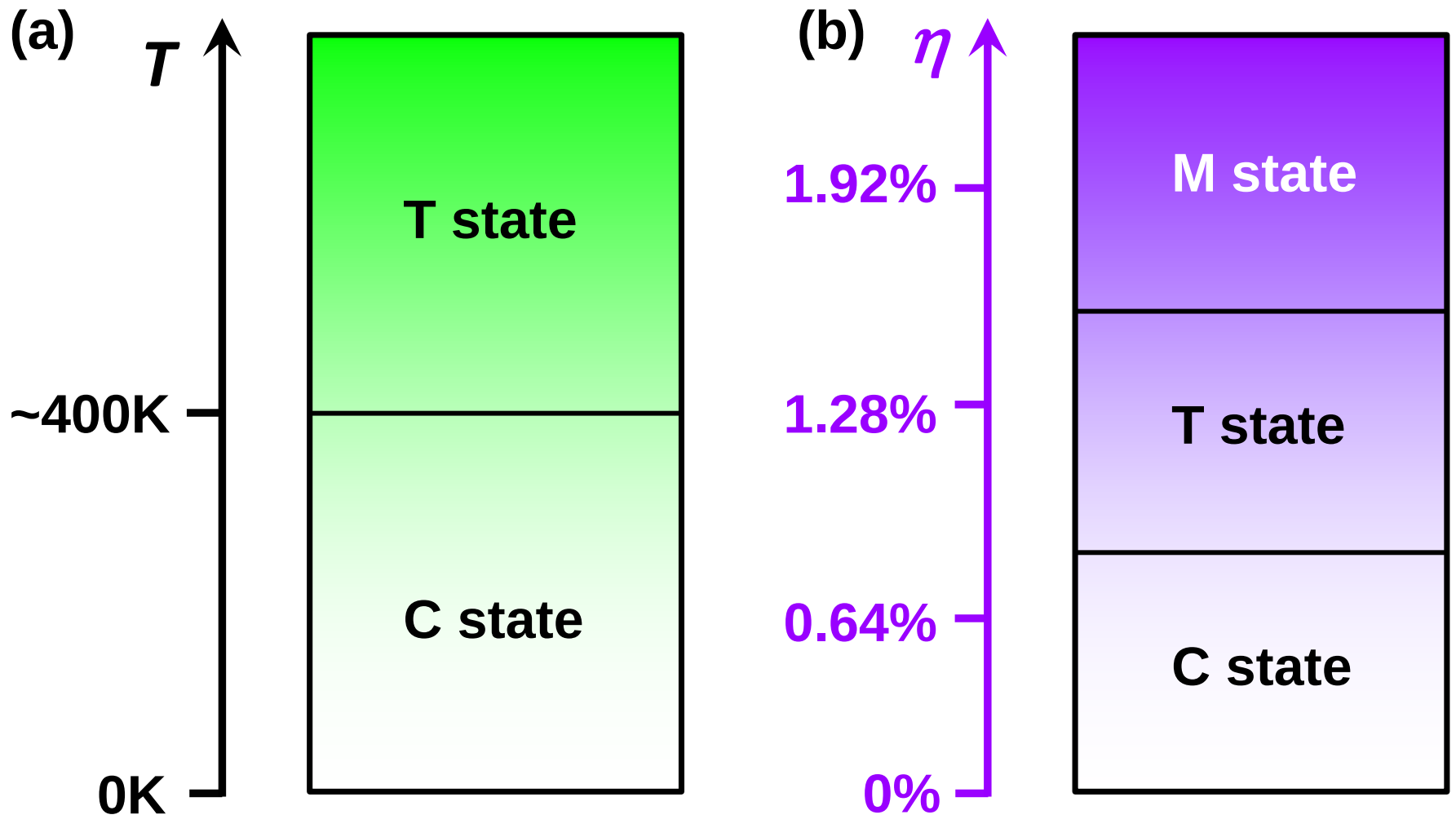

**Figure 5 | Schematic phase diagrams of bulk 1*T*-TaS$_2$.** (a) Under thermal activation and (b) under different laser excitations. The C state of bulk 1*T*-TaS$_2$ has an equilibrium melting point at around 400 K. For $\eta$ < 1.28%, bulk 1*T*-TaS$_2$ remains its low-temperature C state. It transforms to T state at $\eta$=1.28%. The M state emerges at higher laser intensities ($\eta \geq$ 1.92%), which is attributed to the significantly modification of potential surface.

**Table 1. Excitation energy ($E_{ex}$, in unit of meV/atom) for different laser pulses and electronic temperatures ($T_e$) of bulk 1*T*-TaS$_2$**

| Condition | Excitation energy | Condition | Excitation energy |
|---|---|---|---|
| $\eta$ = 0.64% | 40.5 | $T_e$ = 3000K | 52.3 |
| $\eta$ = 1.28% | 87.5 | $T_e$ = 4000K | 92.3 |
| $\eta$ = 1.92% | 141.3 | $T_e$ = 5000K | 153.8 |

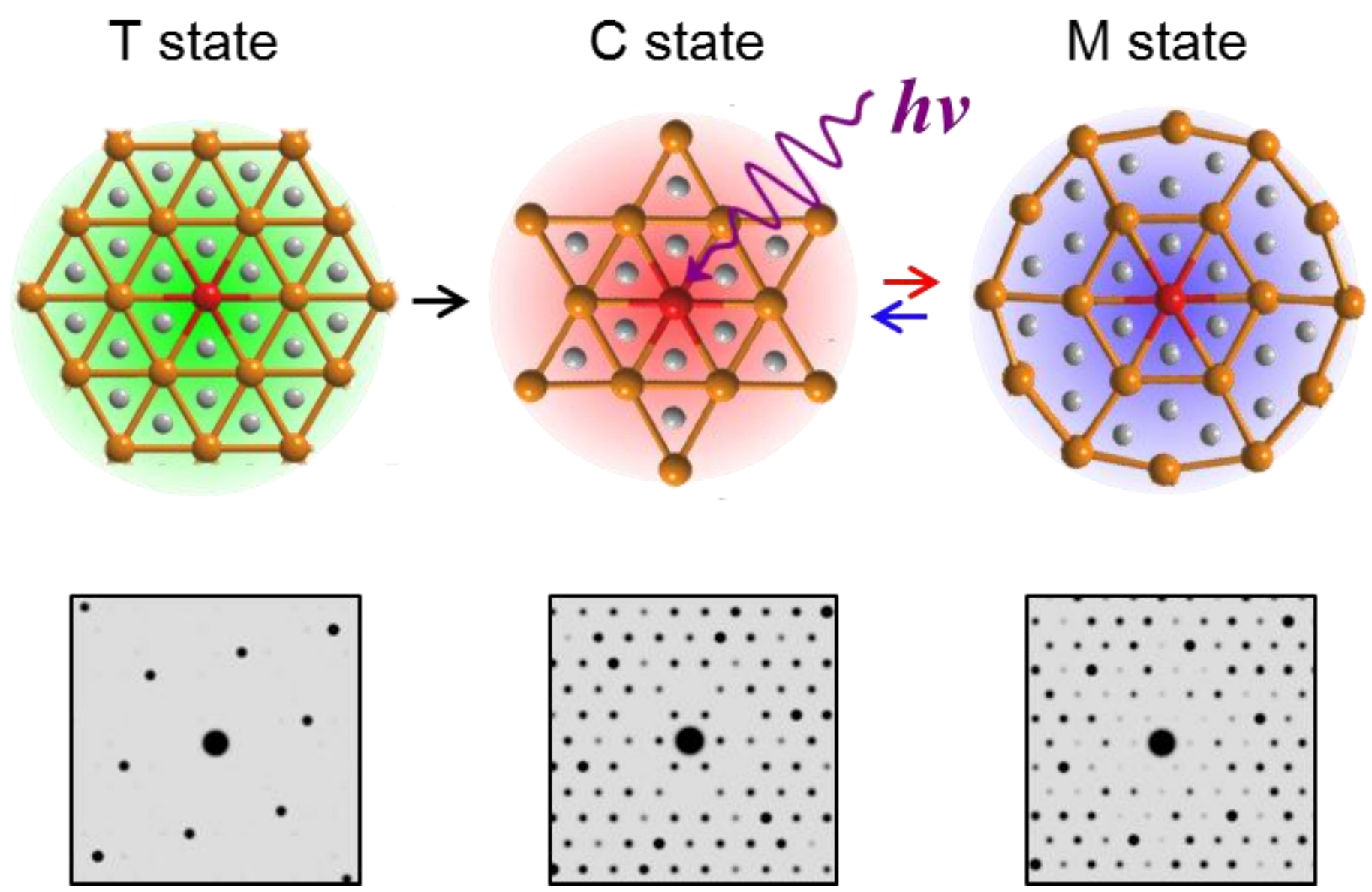

**TOC**